\documentclass{article}
\usepackage{epsfig}
\usepackage{amsmath}
\usepackage{color}
\newtheorem{theorem}{Theorem}[section]
\newtheorem{lemma}{Lemma}[section]
\newtheorem{corollary}{Corollary}[section]

\newcommand{\blackslug}{\penalty 1000\hbox{
    \vrule height 8pt width .4pt\hskip -.4pt
    \vbox{\hrule width 8pt height .4pt\vskip -.4pt
          \vskip 8pt
      \vskip -.4pt\hrule width 8pt height .4pt}
    \hskip -3.9pt
    \vrule height 8pt width .4pt}}

\newenvironment{proof}{\vspace{1mm} \noindent {\sc Proof.}$\;$\rm}{\qed}
\newcommand{\qed}{\hspace*{\fill}\blackslug}
\def\boxit#1{\vbox{\hrule\hbox{\vrule\kern4pt
 \vbox{\kern1pt#1\kern1pt}
\kern2pt\vrule}\hrule}}
\begin{document}

\title{\bf On the Induced Matching Problem in Hamiltonian Bipartite Graphs}

\author{Yinglei Song \\
School of Electronics and Information Science\\
Jiangsu University of Science and Technology\\
Zhenjiang, Jiangsu 212003, China\\
syinglei2013@163.com\\
}
\date{}
\maketitle

\begin{abstract}

\noindent In this paper, we study the parameterized complexity and inapproximability
of the {\sc Induced Matching} problem in hamiltonian bipartite graphs.
We show that, given a hamiltonian cycle in a hamiltonian bipartite graph, the problem is W[1]-hard and cannot be solved in time $n^{o(k^{\frac{1}{2}})}$ unless W[1]=FPT, where $n$ is the number of vertices in the graph. In addition, we show that unless NP=P, the maximum induced matching in a hamiltonian graph cannot be approximated within a ratio of $n^{1-\epsilon}$, where $n$ is the number of vertices in the graph. For a bipartite hamiltonian graph in $n$ vertices, it is NP-hard to approximate its maximum induced matching based on a hamiltonian cycle of the graph within a ratio of $n^{\frac{1}{4}-\epsilon}$, where $n$ is the number of vertices in the graph and $\epsilon$ is any positive constant.
\end{abstract}


\section{Introduction}

Given a graph $G=(V,E)$, an {\it induced matching} is a vertex subset $M \subseteq V$ such that the subgraph
induced by $M$ in $G$ is a matching. The goal of the {\sc Maximum Induced Matching} problem is to find the induced matching of the maximum size in a given graph $G$. Although a maximum matching in a graph can be computed in polynomial
time \cite{galil}, finding the maximum induced matching in a given graph is NP-hard \cite{garey}. The problem remains NP-hard in bipartite graphs \cite{duckworth}. It is therefore highly unlikely to develop algorithms that can solve the problem in bipartite graphs in polynomial time.

In practice, instances of intractable problems are often associated with parameters and it is therefore interesting to study whether practically efficient solutions exist for these problems when their parameters are small positive integers.
Parameterized computation identifies one or a few parameters in some intractable problems and focuses on the development of efficient algorithms for these problems while all parameters are small. Specifically, an NP-hard problem is {\it fixed parameter tractable} if a few parameters $s_1, s_2, \cdots, s_l$ can be identified for the problem and it can be solved in time $O(h(s_1, s_2, \cdots, s_l)n^{c})$, where $n$ is the size of the problem, $h$ is a function that only depends on $s_1, s_2, \cdots, s_l$, and $c$ is a constant independent of all parameters.

A well known example of fixed parameter tractable problems is the {\sc Vertex Cover} problem. The problem is NP-hard and the goal of the problem is to decide whether a graph $G=(V,E)$ contains a vertex cover of size at most $k$ or not. Recent work has developed a parameterized algorithm that improves the upper bound of the problem to $O(1.2852^{k}+k|V|)$ \cite{chen0}. This algorithm has practical values and can be used to efficiently solve the problem when the parameter $k$ is fixed and of a small or moderate value. On the other hand, efficient parameterized solutions are not unknown for some problems and these problems are considered to be parameterized intractable. A hierarchy of parameterized complexity classes have been developed in the theory of parameterized computation to describe the parameterized complexity of these problems. A well known example is the {\sc Independent Set} problem. The goal of the problem is to decide whether a graph contains an independent set of size $k$ or not. The problem has been shown to be W[1]-complete \cite{downey1, downey2}. In other words, all problems in the class of W[1] can be reduced to the {\sc Independent Set} problem in polynomial time. All problems in W[1] are thus fixed parameter tractable if the {\sc Independent Set} problem is fixed parameter tractable. A thorough investigation of topics on parameterized computation and complexity classes can be found in \cite{downey}.

The parameterized induced matching problem is to decide whether a graph contains an induced matching of size $k$ or not, where $k$ is a postive integer parameter. The problem
has been shown to be W[1]-hard for general graphs and thus cannot be solved with an efficient parameterized algorithm \cite{downey1}. However, the problem is fixed parameter tractable for graph families that contain certain structure features. For example, the problem is fixed parameter tractable on planar graphs \cite{moser1}. In addition,
the problem can be solved in polynomial time when the underlying graph is chordal or outerplanar \cite{cameron1,cameron2}.
Structure features of certain graph families can thus lead to efficient solutions for computing a maximum induced matching in graphs in these families.

A {\it bipartite graph} is a graph that does not contain odd cycles. In other words, the vertices in a bipartite graph can be partitioned into two independent sets. Some NP-hard problems become tractable when the underlying graphs are bipartite. For example, a maximum independent set in a bipartite graph can be computed in polynomial time.

A graph is {\it hamiltonian} if it
contains a cycle that contains all of its vertices. A graph is {\it hamiltonian bipartite} if it is both bipartite and
hamiltonian. Hamiltonian bipartite graphs have important applications in wireless communications \cite{kao} and quantumn information processing \cite{child}. Finding
exact solutions for some NP-hard optimization problems in hamiltonian bipartite graphs thus
constitutes an important aspect of algorithmic study. In \cite{moser2}, it is shown that the induced matching problem remains NP-hard and W[1]-hard in bipartite graphs. However, it remains unknown whether the problem can be efficiently solved in hamiltonian bipartite graphs, when a hamiltonian cycle of the graph is available.

Since efficient solutions are unlikely to be available for NP-hard problems, it is interesting to study the possibility to approximately solve these problems in polynomial time. Research in approximate computation focuses on the development of approximate algorithms for some NP-hard problems. In general, a solution generated by an approximate algorithm is guaranteed to be within a ratio of the optimal solution and thus can be practically useful. As a classical example, an approximation ratio of $2.0$ can be achieved for the {\sc Minimum Vertex Cover} problem in polynomial time \cite{Johnson}. However, it is often the case that an intractable problem cannot be approximated within a certain ratio unless NP=P. A well known  example is the {\sc Maximum Independent Set} problem. It has been shown that, unless NP=P, the problem cannot be approximated within a ratio of $n^{1-\epsilon}$ in polynomial time, where  $\epsilon$ is any positive constant and $n$ is the number of vertices in the graph \cite{hastad}. Another example is the {\sc Minimum Vertex Cover} problem. It has been shown that it is NP-hard to approximate the minimum vertex cover in a graph within a ratio of $1.677$ \cite{dinur}. In \cite{raz}, it is shown that unless NP=P, the {\sc Minimum Dominating Set} problem cannot be approximated within a ratio of $c\ln{n}$, where $c$ is some constant independent of $n$ \cite{raz}. Our previous work shows that a similar inapproximability result also holds for the {\sc Minimum Dominating Set} problem when the underlying graph is chordal or near chordal \cite{liu0, liu}.

In spite of its inapproximability in general graphs, the {\sc Minimum Dominating Set} problem can be approximated within a constant ratio of $5$ in planar graphs. The approximate ratio of an intractable problem can thus be significantly improved by certain structure features of the underlying graph. It is therefore natural to ask whether the maximum induced matching problem in a hamiltonian bipartite graph can be approximated within a good ratio based on a hamiltonian cycle of the graph if no efficient solutions are available for the problem.

In this paper, we study the parameterized complexity and inapproximability of the {\sc Induced Matching} problem in
hamiltonian bipartite graphs. We show that, given a
hamiltonian bipartite graph in $n$ vertices and a hamiltonian cycle of the graph, the problem remains W[1]-hard and cannot be solved in time $n^{o(k^{\frac{1}{2}})}$ unless W[1]=FPT. For inapproximability, we show that it is NP-hard to approximate a maximum induced matching in a hamiltonian graph based on a hamiltonian cycle of the graph within a ratio of $n^{1-\epsilon}$, where $n$ is the number of vertices in the graph and $\epsilon$ is any positive constant. For a hamiltonian bipartite graph, its maximum induced matching cannot be approximated based on a hamiltonian cycle of the graph within a ratio of $n^{\frac{1}{4}-\epsilon}$ unless NP=P.

\section{Preliminaries and Notations}
Given a graph $G=(V,E)$, the {\it degree} of a vertex $v \in G$ is the number of vertices that are joined to $v$ with an edge. $G$ is {\it bipartite} if $V$ can be partitioned into
two disjoint subsets $V_1, V_2$ such that
each of $V_1$ and $V_2$ induces an independent set in $G$.
$V_1$ and $V_2$ are the two {\it sides} of $G$. A graph is {\it complete} if any pair of its vertices are joined with an edge.
A bipartite graph is {\it complete} if any pair of vertices
from different sides are joined by an edge.

A {\it hamiltonian path} in a graph is a path that contains each vertex in the graph. A graph is {\it hamiltonian} if it
contains a cycle that includes all its vertices and such a cycle is a {\it hamiltonian cycle}. A graph is a {\it hamiltonian bipartite} graph if it is both bipartite
and hamiltonian. An {\it induced matching} in $G$ is a vertex
subset $U \subseteq V$ such that the subgraph induced by
$U$ in $G$ is a matching. The {\sc Induced Matching} problem
is to decide whether a given graph $G$ contains an
induced matching of size $k$ or not.

Given a graph $G=(V,E)$, a subset of vertices $v_1, v_2, \cdots, v_k$ in $V$ form a {\it clique} if any pair of vertices in the subset are joined by an edge in $G$. The {\sc Clique} problem is to decide whether a given graph contains a clique of size $k$ or not, where $k$ is a positive integer. A well known fact is that the problem is W[1]-hard \cite{downey}. In other words, the problem cannot be solved in time $O(f(k)n^{c})$ unless the complexity class of W[1] collapses into FPT, where $f$ is a function of $k$, $n$ is the number of vertices in the graph and $c$ is a contant independent of $n$ or $k$.

We consider a variant of the {\sc Clique} problem. Given a graph $G=(V,E)$, the goal of this variant is to determine whether $G$ contains a clqiue of size $2k+1$ or not, where $k$ is a positive integer. We denote the variant by $(2k+1)$-{\sc Clique} problem in the rest of the paper and it can be shown that the problem remains W[1]-hard.

\begin{lemma}
\rm
\label{lm1}
The $(2k+1)$-{\sc Clique} problem is W[1]-hard and it cannot be solved in time $n^{o(k)}$ unless W[1]=FPT, where $n$ is the number of vertices in the graph.

\begin{proof}
We construct a polynomial time reduction from the {\sc Clique} problem to the $(2k+1)$-{\sc Clique} problem. Given a graph $G=(V,E)$, we construct a graph $H$ such that $G$ contains a clique of size $k$ if and only if $H$ contains a clique of $2k+1$.

To complete the construction, we create two identical copies $G_1$, $G_2$ of $G$. In other words, both $G_1$ and $G_2$ are isomorphic to $G$. Any two vertices from different copies of $G$ are joined with an edge. We then generate one additional vertex $u$ and join each vertex in $G_1$, $G_2$ to $u$ with an edge. We denote the resulting graph by $H$.

If $G$ contains a clique of size $k$, $H$ also contains a clique of size $2k+1$. Specifically, since $G_1$, $G_2$ are both isomorphic to $G$, each of them contains $k$ vertices that are connected into a clique in $H$. It is not difficult to see that these $2k$ vertices and $u$ together form a clique of size $2k+1$ in $H$.

We now assume $H$ contains a clique $C$ of size $2k+1$, we use $C_1$ and $C_2$ to denote $C \cap G_1$ and $C \cap G_2$ respectively. Since $u$ is the only vertex that is not included in $G_1$ or $G_2$, we immediately obtain
\begin{equation}
    |C_1|+|C_2| \geq 2k
\end{equation}

The above inequality implies that one of $C_1$ and $C_2$ must contain at least $k$ vertices and these vertices are connected into a clique. Since $G_1$, $G_2$ are both isomorphic to $G$, this implies that $G$ contains a clique of size at least $k$. The $(2k+1)$-{\sc Clique} problem is thus also W[1]-hard.

We assume that there exists an algorithm $A$ that can solve the $(2k+1)$-{\sc Clique} problem in time $n^{o(k)}$, where $n$ is the number of vertices in the graph. Given a graph $G=(V,E)$ and a positive integer parameter $k$, we can use the following algorithm to solve the {\sc Clique} problem.
\begin{enumerate}
\item{Construct a graph $H$ from $G$ based on the reduction that has been described above;}
\item{apply $A$ to $H$ to determine whether it contains a clique of size $2k+1$ or not;}
\item{return ``yes'' if $A$ returns ``yes'', otherwise return ``no''.}
\end{enumerate}

It is straightforward to see that the above algorithm solves the {\sc Clique} problem in time $|V|^{o(k)}$ since $H$ contains $2|V|+1$ vertices. However, it has been shown in \cite{chen} that the {\sc Clique} problem cannot be solved in time $|V|^{o(k)}$ unless W[1]=FPT. Such an algorithm thus does not exist for the $(2k+1)$-{\sc Clique} problem unless W[1]=FPT.
\end{proof}
\end{lemma}

Given a graph $G=(V,E)$, a {\it walk} in $G$ is a sequence of edges $e_1, e_2, \cdots, e_n$ such that $e_i$ and $e_{i+1}$ share an endpoint, where $1 \leq i <n$. A walk is an {\it eulerian path} if each edge in $G$ is visited for exactly once. The {\it starting vertex} of an eulerian path $e_1, e_2, \cdots, e_n$ is the endpoint of $e_1$ that is not shared with $e_2$ and its {\it ending vertex} is the endpoint of $e_{n}$ that is not shared with $e_{n-1}$. An eulerian path is an {\it eulerian circuit} if its starting and ending vertices are the same vertex. $G$ is {\it eulerian} if it contains one eulerian circuit. In \cite{biggs}, it is shown that a graph is {\it eulerian} if the degree of each vertex in the graph is even. It is shown in \cite{fleury} that an eulerian circuit, if exists, can be computed in polynomial time.

\begin{lemma}
\rm
\label{lm2}
A complete graph on an odd number of vertices is eulerian and such a path can be computed in polynomial time.

\begin{proof}
The degree of each vertex in a complete graph on $n$ vertices is $n-1$. $n-1$ is even when $n$ is odd. The lemma follows from the results in \cite{biggs} and \cite{fleury}.
\end{proof}
\end{lemma}

Given a graph $G=(V,E)$, an {\it edge graph} $H$ of $G$ can be constructed by representing each edge in $G$ by a distinct vertex in $H$ and two vertices in $H$ are joined by an edge if their corresponding edges in $G$ share an endpoint.

\begin{lemma}
\rm
\label{lm3}
Given a complete graph $G$ on an odd number of vertices, its edge graph $H$ is hamiltonian.

\begin{proof}
We denote the number of edges in $G$ by $n$. From Lemma \ref{lm2}, there exists an eulerian circuit $e_1, e_2, \cdots, e_n$ in $G$. We use $u_i$ to denote the vertex that corresponds to $e_i$ in $H$, where $1 \leq i \leq n$. It is clear that $u_1, u_2, \cdots, u_n$ is in fact a hamiltonian path in $H$. In addition, since $e_1$ and $e_n$ share an endpoint, $u_1$ and $u_n$ are joined by an edge in $H$. $H$ is thus hamiltonian.
\end{proof}
\end{lemma}

\begin{lemma}
\rm
\label{lm4}
Given a complete bipartite graph $G=(V_1 \cup V_2, E)$, where
$V_1$ and $V_2$ are the two sides of $G$ and $|V_1|=|V_2|$, for any two vertices $u$, $v$ such that $u \in V_1$ and $v \in V_2$, there exists a hamiltonian path that starts with $u$ and ends with $v$. Such a hamiltonian path can be computed in polynomial time.

\begin{proof}
We denote $|V_1|$ and $|V_2|$ by $n$. We consider the vertices in $V_1-\{u\}$ and $V_2-\{v\}$, since $|V_1|=|V_2|$, each vertex in $V_2-v$ can be arbitrarily paired with a distinct vertex in $V_1-u$. All vertices in $V_1-\{u\}$ and $V_2-\{v\}$ can thus be partitioned into $n-1$ disjoint vertex pairs, we use $(v_1, u_1), (v_2, u_2), \cdots, (v_{n-1}, u_{n-1})$ to denote them, where $v_i \in V_2-\{v\}$ and $u_i \in V_1-\{u\}$ ($1 \leq i \leq n-1$). Since $G$ is complete bipartite, $u, v_1, u_1, v_2, u_2, \cdots, v_{n-1}, u_{n-1}, v$ is a hamiltonian path in $G$. It is also clear that the path can be obtained in polynomial time.
\end{proof}
\end{lemma}

\section{Parameterized Complexity and Lower Bound}

The {\sc Induced Matching} problem has been shown to be NP-complete in general graphs and bipartite graphs. In addition, the parameterized induced matching problem is
shown to be W[1]-hard in general graphs. In \cite{moser2}, a reduction from the {\sc Irredundant Set} problem \cite{downey3} is constructed to show that the problem remains W[1]-hard in bipartite graphs. In this section, we construct a reduction from the $(2k+1)$-{\sc Clique} problem to show that the problem remains W[1]-hard in hamiltonian bipartite graphs.

\begin{figure}[!tcp]
\begin{center}
\includegraphics[width=10.0cm, height=7.8cm]{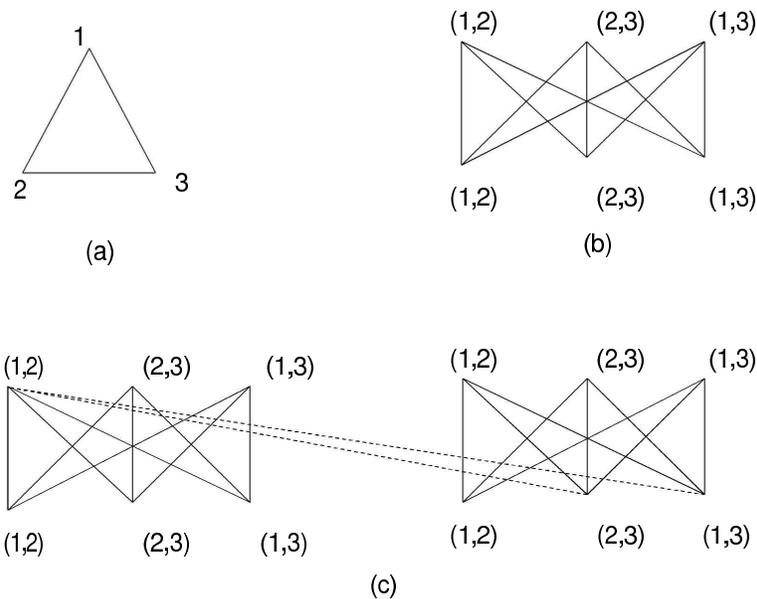}
\caption{A gadget constructed from a simple graph.}
\label{fig1}
\end{center}
\end{figure}

\begin{theorem}
\label{th1}
\rm
Given a hamiltonian bipartite graph and a hamiltonian cycle in the graph, it is W[1]-hard to determine whether the graph contains an induced matching of size $p$ or not, where $p$ is a positive integer parameter.

\begin{proof}
We construct a polynomial-time reduction from the $(2k+1)$-{\sc Clique} problem, the goal of this problem is to
decide whether a given graph $G=(V,E)$ contains a clique of size $2k+1$ or not. From Lemma \ref{lm1}, the problem is
W[1]-hard. Given a graph $G=(V,E)$ we construct a hamiltonian bipartite graph $H$ such that $G$ contains a clique of
size $l=2k+1$ if and only if $H$ contains an induced matching of size $O(k^2)$. Without loss of generality, we assume $G$ contains at least $7$ vertices and $3$ edges.

To construct the graph $H$, we create $\frac{1}{2}l(l-1)$ graph
gadgets, each gadget is designed to select an edge in the
graph, the $\frac{1}{2}l(l-1)$ selected edges then form a clique of size $l$ in $G$. Each gadget contains three complete bipartite subgraphs, such a complete bipartite graph is a {\it bipartite unit}. Each bipartite unit in the gadget
contains $|E|$ vertices, and each vertex represents an edge in $G$. The three bipartite units are connected into a larger bipartite graph. In particular, the vertices in all three bipartite units are grouped into two sides, two vertices from two different bipartite units are joined by an edge if they are in different sides and represent
different edges in $G$. The vertices in two sides are included in two disjoint vertex subsets $S_1$ and $S_2$ respectively.

Figure \ref{fig1} shows a gadget
constructed from a simple graph. In the figure, (a) shows a graph that contains vertices 1, 2 and 3;
(b) is a bipartite unit constructed from the edges in the graph in (a); (c) shows a gadget constructed by connecting
two bipartite units into a larger graph. Solid lines in the figure are the edges in the bipartite units and
dashed lines represent edges that connect vertices from different bipartite units.
To make the figure clear, only the edges that join the vertex$(1,2)$ in one bipartite unit to vertices in the other one are shown in the figure.

Each graph gadget is associated with an integer pair $(k_1, k_2)$ such
that $1 \leq k_1 < k_2 \leq l$, the pair represents the
edge that joins the vertices for $k_1$ and $k_2$ in a clique of size $l$ in $G$. One of the bipartite unit in a graph gadget $(k_1, k_2)$ is
dedicated to select a vertex in $G$ for $k_1$ in the clique and is called {\it $k_1$ unit}. Another
one is dedicated to select a vertex for $k_2$ and is called
{\it $k_2$ unit}. The remaining one is designed to guarantee that both bipartite units select the same edge.

We construct a graph $D$ to describe the relationships among different graph gadgets. Each graph gadget is represented by a vertex in $D$ and two vertices are joined by an edge if the corresponding gadgets share an integer in their integer pairs. For example, vertices that represent the graph gadgets with integer pairs $(k_1, k_2)$ and $(k_1, k_3)$ are joined by an edge in $D$ since their integer pairs share an integer $k_1$.

It is clear that $D$ is isomorphic to the edge graph of a complete graph on $l$ vertices. From Lemma \ref{lm3} $D$ is a hamiltonian graph since $l=2k+1$ is an odd integer. There exists a hamiltonian cycle $C=g_1, g_2, \cdots, g_{k(2k+1)}, g_1$ in $D$, where the gadgets represented by any two vertices consecutive in $C$ share an integer in their integer pairs. It is also clear from Lemma \ref{lm3} that $C$ can be computed in polynomial time.

Two graph gadgets are {\it consecutive} if their corresponding vertices are consecutive in $C$. Based on $C$, each pair of gadgets that are consecutive in $C$ are connected by a {\it connector}. We next describe how
a connector is constructed. Similar to the gadgets we have constructed to select edges, a connector contains $3$ bipartite units, each bipartite unit in a connector contains $|V|$ vertices, and each vertex represents a vertex in $G$.
The three bipartite units in a connector are connected into a larger bipartite graph. The vertices in all three bipartite units are grouped into two sides. Vertices in one side are included in $S_1$ and those in the other side are included in $S_2$. Two vertices in different bipartite units
are joined by an edge if one of them is in $S_1$ and the other is in $S_2$ and they represent different vertices in $G$.

\begin{figure}[!tcp]
\begin{center}
\includegraphics[width=10.0cm, height=7.8cm]{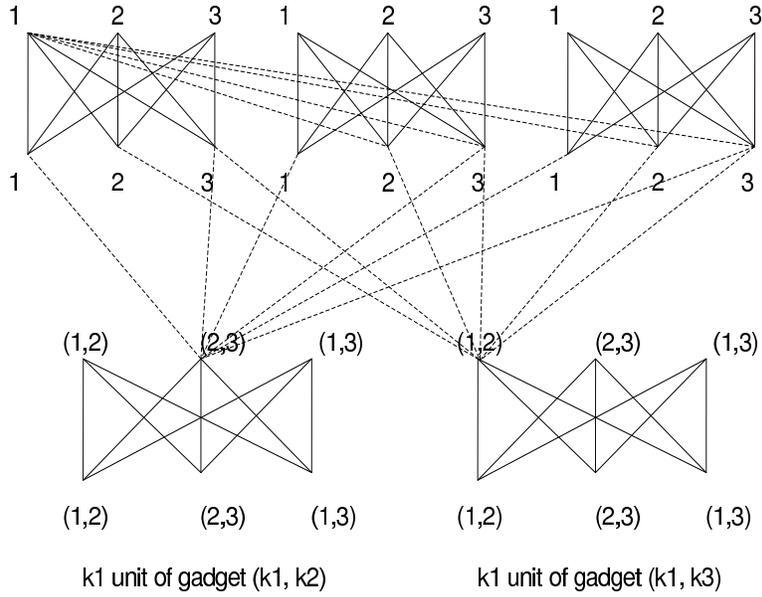}
\caption{A connector and the edges that connect it to two other
gadgets.}
\label{fig2}
\end{center}
\end{figure}

Given two consecutive graph gadgets $(k_1,k_2)$ and $(k_1, k_3)$, we connect the $k_1$ units of the two gadgets with a connector. The two $k_1$ units are connected to the three bipartite units in a connector into a larger bipartite graph. Specifically, for a given vertex $u$ in a $k_1$ unit that represents edge $(a,b)$ in $G$, we join each vertex from the other side that do not represent $a$ in the connector to $u$ with an edge. A connector is {\it for $k_1$} if it connects two graph gadgets whose integer pairs both contain $k_1$.

Figure \ref{fig2} shows an example
of a connector and the edges that connect the connector to the two consecutive graph gadgets $(k_1, k_2)$ and $(k_1, k_3)$.
Solid lines in the figure are the edges in the bipartite
units and dashed lines represent the edges that join vertices from different bipartite units. To make the figure clear, only part of the edges are shown.

As the last step of the construction, we connect all connectors for each $t$, where $1 \leq t \leq l$, into a larger bipartite graph. For a given $t$, we use $N(t)$ to denote all vertices in all connectors for $t$. An edge is created to join two vertices $c \in N(t)$ and $d \in N(t)$ if one of them is in $S_1$, the other one is in $S_2$ and they represent different vertices in $G$. The subgraph induced by vertices in $N(t)$ is the {\it connector group} for $t$.

\begin{lemma}
\rm
\label{lm5}
$H$ is a hamiltonian bipartite graph and a hamiltonian cycle in $H$ can be computed in polynomial time.

\begin{proof}
From the construction of $H$, any vertex in $H$ is either in $S_1$ or $S_2$ and any edge in $H$ is between a vertex in $S_1$ and a vertex in $S_2$. $H$ thus must be bipartite. We then show that a hamiltonian cycle in $H$ can be constructed based on $C$. We use $g_1, g_2, g_3, \cdots, g_{k(2k+1)}$ to denote the graph gadgets that correspond to the vertices in $D$ along $C$. Recall that $C$ is constructed from an eulerian circuit in a complete graph, one of the integers in the integer pair of $g_i$ is shared with that of $g_{i+1}$ and the other one is shared with that of $g_{i-1}$ for $1< i< k(2k+1)$. For $g_1$, the integers in its integer pair are shared with $g_2$ and $g_{k(2k+1)}$ respectively. Similarly, the integers in the integer pair of $g_{k(2k+1)}$ are shared with $g_1$ and $g_{k(2k+1)-1}$ respectively.

We denote the integer shared between the integer pairs of $g_{i}$ and $g_{i+1}$ with $b(i)$ for $1 \leq i < k(2k+1)$ and use $b(k(2k+1))$ to denote the integer shared between the integer pairs of $g_{k(2k+1)}$ and $g_1$. It is clear that $g_{i}$ contains a bipartite unit for $b(i-1)$ and another one for $b(i)$ for $1 < i \leq k(2k+1)$. For each such $i$, we arbitrarily choose a vertex $s_i$ that are in both $S_1$ and the unit for $b(i)$ in $g_i$ and a vertex $e_i$ that are in both $S_2$ and the unit for $b(i+1)$ in $g_i$. From Lemma \ref{lm4}, since $G$ contains at least $3$ edges we can find a path $p_i$ that starts with $s_i$, visits every vertex in $g_{i}$ and ends with $e_i$. Similarly, for $g_1$, we arbitrarily choose $s_1$ in both $S_1$ and the unit for $b(k(2k+1))$ and $e_1$ in both $S_2$ and the unit for $b(1)$ and there exists a path $p_1$ that starts with $s_1$, visits every vertex in $g_1$ and ends with $e_1$.

For $1 \leq i < k(2k+1)$, both $e_i$ and $s_{i+1}$ are connected to a connector. Note that since $e_i \in S_2$, $s_{i+1} \in S_1$ and $G$ contains at least $7$ vertices, we use Lemma \ref{lm4} again and there exists a path $q_i$ that starts with $e_i$, visits every vertex in the connector and ends with $s_{i+1}$. For the same reason, we can find a path $q_{k(2k+1)}$ that starts with $e_{2k(2k+1)}$, visits every vertex in the connector between $g_{k(2k+1)}$ and $g_{1}$ and ends with $s_1$. It is not difficult to see that $p_1, q_1, p_2, q_2, \cdots, p_{k(2k+1), q_{k(2k+1)}}$ form a hamiltonian cycle in $H$. $H$ is thus hamiltonian bipartite. From Lemma \ref{lm4}, the cycle can be obtained in polynomial time.
\end{proof}
\end{lemma}

\begin{lemma}
\rm
\label{lm6}
$G$ contains a clique of size $2k+1$ if and only if $H$ contains an induced matching of size $6k(2k+1)$.

\begin{proof}
First, if $G$ contains a clique of size $l$, we denote the vertices in the clique with $a_1, a_2, \cdots, a_{l}$ and they correspond to $1, 2, 3, \cdots, l$ in the clique. For gadget $(k_1, k_2)$, we select the three edges that represent edge $(a_{k_1}, a_{k_2})$ in the three bipartite units in the gadget. There are in total $3k(2k+1)$ such edges. In addition to these edges, for each connector, we select the $3$ edges that represent the vertex shared by the two gadgets associated with the connector. For example, for the connector that connects gadgets $(k_1, k_2)$ and $(k_1, k_3)$, we choose the $3$ edges that represent vertex $a_{k_1}$ in the $3$ bipartite units in the connector.
Since there are $k(2k+1)$ connectors in total, we have $3k(2k+1)$ edges selected in connectors. The total number of edges selected is thus $6k(2k+1)$. From the construction of $H$, it is not difficult to see that the selected edges together form an induced matching.

Next, we show that if $H$ contains an induced matching $M$ of size $6k(2k+1)$, $G$ contains a clique of size $2k+1$. We first show that the subgraph induced by vertices in a gadget may contain at most three edges in an induced matching of $H$. To show this, we assume there exists a gadget $T$ that contains four edges in $M$. Since $T$ contains three bipartite units, there exists two vertices $c_1 \in M$ and $c_2 \in M$ such that $c_1$ and $c_2$ are in the same side of a bipartite unit in $T$. Without loss of generality, we assume that $c_1$ and $c_2$ are both in $S_1$. Since $T$ contains four edges in $M$, it contains a vertex $d$ such that $d \in S_2$, $d \in M$ and $d$ is not matched to $c_1$ or $c_2$ in $M$. However, $d$ is joined to at least one of $c_1$ and $c_2$ by an edge since $c_1$ and $c_2$ represent two different edges in $G$. This contradicts the fact that $M$ is an induced matching. $T$ thus cannot contain more than three edges in $M$. On the other hand, $M$ may contain three edges in a gadget if each bipartite unit in the gadget contains one edge in $M$ and all three edges represent the same edge in $G$.

We consider the case where $M$ contains three edges in a gadget $T$. We partition the edges in $T$ into two sets $D_1$ and $D_2$ such that $D_1$ contains the edges that join vertices in the same bipartite unit and $D_2$ contains the edges that join vertices from different bipartite units. We show
that if $M$ contains three edges in $T$, all three must come from $D_1$. First, if only one edge is from $D_2$, there must exist three vertices in $M$ such that one of them is in one side of a bipartite unit and the other two are in the other side of the same bipartite unit. This is contradictory to the fact that $M$ is an induced matching since a bipartite unit is a complete bipartite graph. If two edges are from $D_2$, there exists two vertices $c_1 \in M$ and $c_2 \in M$ such that $c_1$ and $c_2$ are in the same side of a bipartite unit in $T$. Since $T$ contains three edges in $M$, there exists a vertex $d$ such that $d \in M$, $d$ is not matched to $c_1$ or $c_2$ in $M$ and $d$ is in the other side of $T$. Again, $d$ must be joined to at least one of $c_1$ and $c_2$ by an edge since they represent different edges in $G$. This contradicts the fact that $M$ is an induced matching. All three edges thus must be in $D_1$ and they must represent the same edge in $G$. We thus have shown that $M$ can contain at most three edges from a gadget and if $M$ contains three edges in a gadget, each of them represents an edge in $G$ and the three edges must represent the same edge in $G$.

For each $1 \leq t \leq l$, we consider the bipartite units in the connector group for $t$. From the construction, the connector group contains $3k$ bipartite units in total. An $m$ {\it partial connector group} for $t$ is the subgraph induced by vertices in $m$ bipartite units in the connector group for $t$, where $m$ is a positive integer and $1 \leq m \leq 3k$. We show that $M$ contains at most $m$ edges in an $m$ partial connector group. The claim trivially holds when $m=1$. We now consider the case where $m>1$. Indeed, we assume $M$ contains $m+1$ edges in an $m$ partial connector group for $t$. Since the $m$ partial connector group contains at most $m$ bipartite units, there exists a bipartite unit that contains two vertices $c_1 \in M$, $c_2 \in M$ such that $c_1$ and $c_2$ are both in the same side of the bipartite unit. Since $M$ contains $m+1$ edges in the $m$ partial connector group, there exists a vertex $d$ such that $d \in M$, $d$ is not matched to $c_1$ or $c_2$ in $M$ and $d$ is in the other side of the $m$ partial connector group. $d$ must be joined to one of $c_1$ and $c_2$ since they represent different vertices in $G$. This contradicts the fact that $M$ is an induced matching. $M$ thus contains at most $m$ edges in an $m$ partial connector group for $t$.

We then show that if $M$ contains $3k$ edges in the connector group for $t$, each of these edges must be completely contained in one bipartite unit of the connector group and represents the same vertex in $G$. We assume this is not the case and there exists an edge $(c_1, c_2)$ in $M$ such that $c_1 \in S_1$ is in bipartite unit $b_1$ and $c_2 \in S_2$ is in bipartite unit $b_2$. Since both $b_1$ and $b_2$ are complete bipartite, $M$ does not contain vertices in the $S_2$ side of $b_1$ or the $S_1$ side of $b_2$. $M$ must contain another vertex in the $S_1$ side of $b_1$ or the $S_2$ side of $b_2$ since otherwise the remaining $3k-2$ bipartite units form a $3k-2$ partial connector group and $M$ contains at most $3k-2$ edges in this partial connector group. The total number of edges $M$ contains in the connector group is thus at most $3k-1$, which contradicts the fact that $M$ contains $3k$ edges in the connector group. $M$ thus must contain two different vertices $d_1$ and $d_2$ in one side of $b_1$ or $b_2$. Since $3k>2$, $M$ must contain a vertex $d$ such that $d$ is on the other side of the connector group and is not matched to $d_1$ or $d_2$. Again, $d$ must be joined to at least one of $d_1$ and $d_2$ since they represent different vertices in $G$. This contradicts the fact that $M$ is an induced matching. Such an edge thus does not exist in $M$. Each edge in $M$ must be completely contained in one bipartite unit. Since $3k \geq 3$, it is straightforward to see that these edges must represent the same vertex in $G$.

An edge is an {\it inner edge} if it joins two vertices that are both in the same graph gadget or connector group. An edge is a {\it boundary edge} if it is not an inner edge. A boundary edge is {\it attached} to the connector group $t$ if one of its endpoints is in the connector group for $t$, where $1 \leq t \leq l$. Given an induced matching $M$ in $H$,  a gadget with integer pair $(k_1, k_2)$ is {\it bad in $k_1$} if $M$ contains a boundary edge $e$ such that one of the endpoints of $e$ is in the $k_1$ unit of the gadget. A gadget with integer pair $(k_1, k_2)$ is {\it completely bad} if it is bad in both $k_1$ and $k_2$. A gadget with integer pair $(k_1, k_2)$ is {\it bad in one side} if it
is bad in either $k_1$ or $k_2$ but not completely bad. A gadget is {\it good} if it is neither completely bad nor bad in one side.

We show that $M$ contains at most $2$ edges in a gadget that is not good. We denote the integer pairs of the gadget with $(k_1, k_2)$. If $M$ contains $3$ edges in the gadget, each edge of the three is completely contained in one bipartite unit of the gadget. However, since the gadget is not good, a different vertex in the $k_1$ unit or the $k_2$ unit is also included in $M$. This contradicts the fact $M$ is an induced matching since both $k_1$ and $k_2$ units are complete bipartite. $M$ thus contains at most $2$ edges in a gadget that is not good.

For any $t$ such that $1 \leq t \leq l$, we consider all boundary edges that are in $M$ and attached to the connector group for $t$. First, we observe that each bipartite unit in the connector group for $t$ contains at most two vertices that are the endpoints of boundary edges in $M$. We assume there exists a bipartite unit that contains three such vertices $c_1$, $c_2$, $c_3$ and $c_1$ is matched in $M$ to a vertex $e$ that represents an edge in $G$. It is clear that $c_1$, $c_2$, $c_3$ must be in the same side of the bipartite unit. In addition, since $c_2$ and $c_3$ represent two different vertices in $G$, one of them must be joined to $e$ by an edge and this contradicts the fact that $M$ is an induced matching. Such a bipartite unit thus does not exist.

Given an induced matching $M$, a bipartite unit in the connector group for $t$ is {\it full} if it contains two vertices that are the endpoints of two boundary edges. A bipartite unit is {\it empty} if it does not contain any vertices in $M$ and is {\it lonely} if it is neither {\it full} nor {\it empty}. We now consider a full bipartite unit $b_1$ that contains two vertices $c_1$ and $c_2$ that are the endpoints of two boundary edges in $M$. We assume $c_1$ and $c_2$ are matched in $M$ to vertices $e_1$ and $e_2$ that represent two edges in $G$. We use $b_2$ and $b_3$ to denote the other two bipartite units that are in the same connector as $b_1$. We show that both $b_2$ and $b_3$ must be empty. Without loss of generality, we assume that both $c_1$ and $c_2$ are in $S_1$ side of $b_1$. We first show that $M$ does not contain a vertex in the $S_2$ side of $b_2$ or $b_3$. We assume such a vertex $d$ exists, $d$ must be joined to at least one of $c_1$ and $c_2$ by an edge in $H$ since they represent different vertices in $G$. This contradicts the fact that $M$ is an induced matching. Such a vertex thus does not exist.

We then show that $M$ does not contain a vertex in the $S_1$ side of $b_2$ or $b_3$. We assume there exists such a vertex $d$. Since $e_1$ is not joined to $c_2$ by an edge in $H$, the vertex represented by $c_2$ must be the corresponding endpoint of the edge represented by $e_1$. Since both $d$ and $e_1$ are in $M$ and $d$ is not matched to $e_1$ in $M$, $d$ must also represents the vertex represented by $c_2$. However, since $c_2$ is joined to $e_2$ by an edge in $H$, $d$ is also joined to $e_2$ by an edge in $H$. This contradicts the fact that $M$ is an induced matching. Such a vertex thus does not exist in the $S_1$ side of $b_2$ or $b_3$. $M$ thus does not contain any vertices in $b_2$ or $b_3$.

It is also clear that $b_1$ does not contain a third vertex that is the endpoint of an inner edge in $M$. To show this, we assume such a vertex $c_3$ exists, $c_3$ must be in the same side as that of $c_1$ and $c_2$. We assume $c_3$ is matched to $d$ in $M$, $d$ must be joined to at least one of $c_1$ and $c_2$ by an edge in $H$ since they represent different vertices in $G$. This contradicts the fact that $M$ is an induced matching.

Given an induced matching $M$ in $H$, a lonely bipartite unit in the connector group for $t$ is {\it occupied} if $M$ contains a boundary edge $e$ such that one of the endpoints of $e$ is included in the bipartite unit. A lonely bipartite unit in the connector group is {\it unoccupied} if it is not occupied. We assume that $M$ contains $s$ boundary edges that are attached to the connector group for $t$ and $2q$ of these $s$ edges are associated with $q$ full bipartite units. Each of the remaining $s-2q$ edges is then associated with a distinct occupied bipartite unit. Recall that there are at least $2q$ empty bipartite units. We immediately obtain that the number of unoccupied bipartite units in the connector group for $t$ is at most $3k-q-(s-2q)-2q=3k-s-q$ if $M$ includes $s$ boundary edges that are attached to the connector group.

Given an induced matching $M$ in $H$, an inner edge $e$ in the connector group for $t$ is {\it dangling} if $e \in M$ and one of the endpoints of $e$ is in an occupied bipartite unit. It is clear that the connector group for $t$ does not contain a dangling edge if it does not have unoccupied bipartite units. In addition, We observe that if the connector group for $t$ contains one dangling edge $f$, $f$ is the only inner edge that $M$ contains in the connector group. Indeed, since one of the endpoints of $f$ is in an occupied bipartite unit $b'$, $M$ must contain two vertices $c_1$ and $c_2$ that are in the same side of $b'$. If $M$ contains another inner edge in the connector group, there exists a vertex $d$ such that $d \in M$ and $d$ is not matched to $c_1$ or $c_2$ in $M$. Again, $d$ must be joined to one of $c_1$ and $c_2$ since they represent different vertices in $G$. This contradicts the fact that $M$ is an induced matching. Such an inner edge thus does not exist.

An edge in $H$ {\it belongs to} the connector group for $t$ if one of its endpoints is in the connector group. It is clear from the above reasoning that the number of edges that belong to the connector group for $t$ in $M$ is at most $3k-q$ if it contains a dangling edge.

We now consider the case where the connector group for $t$ does not contain a dangling edge. Since the number of unoccupied bipartite units is $3k-s-q$ and the connector group does not contain dangling edges, all the remaining inner edges in $M$ and the connector group must be completely contained in the $3k-s-q$ partial connector group formed by these unoccupied bipartite units. From the results we have shown for a partial connector group, $M$ contains at most $3k-s-q$ edges in such a partial connector group and the number of edges that belong to the connector group is thus again at most $3k-q$.

We are now ready to show that $G$ must contain a clique of size $2k+1$. We assume $M$ contains $w$ gadgets that are not good and use $e(M)$ to denote the number of edges in $M$. It is not difficult to see that the following inequality holds for $M$.
\begin{equation}
    e(M) \leq 3(k(2k+1)-w)+2w+\sum_{t=1}^{l}B(t)
\end{equation}
where $B(t)$ is the number of edges in $M$ that belong to the connector group for $t$. Since we have shown above that $B(t) \leq 3k-q \leq 3k$. We immediately obtain
\begin{equation}
    e(M) \leq 6k(2k+1)-w
\end{equation}
On the other hand, $M$ contains $6k(2k+1)$ edges. It is thus clear that $w$ must be $0$, which implies that each graph gadget in $H$ must be good. $M$ thus does not include boundary edges. Since We have shown above that $M$ contains at most $3$ inner edges in a graph gadget and $3k$ inner edges in a connector group, we can conclude that $M$ contains exactly $3$ edges in each graph gadget and $3k$ inner edges in each connector group.

As we have shown above, the $3k$ edges in each connector group represent a vertex in $G$. We can thus obtain $l=2k+1$ vertices $a_1, a_2, \cdots, a_{l}$ from the edges that are in $M$ and all connector groups. In addition, each graph gadget selects an edge in $G$. Since $M$ is an induced matching, these edges connect $a_1, a_2, \cdots, a_{l}$ into a clique of size $l=2k+1$. $G$ thus must contain a clique of size $2k+1$.
\end{proof}
\end{lemma}

From Lemma \ref{lm4}, $H$ is hamiltonian bipartite and a hamiltonian cycle $P$ can be obtained in polynomial time. From Lemma \ref{lm5}, $G$ contains a clique of size $2k+1$ if and only if $H$ contains an induced matching of size $6k(2k+1)$. Since $H$ can be constructed in polynomial time, the induced matching problem is W[1]-hard in a hamiltonian bipartite graph, even when a hamiltonian cycle in the graph is available.
\end{proof}
\end{theorem}

Based on the proof of Theorem \ref{th1}, we can obtain a
parameterized lower bound of induced matching problem in hamiltonian bipartite graphs. Since we have shown in Lemma \ref{lm1} that the $(2k+1)$-{\sc Clique} problem cannot be solved in time $n^{o(k)}$ unless W[1]=FPT, we can immediately obtain the following theorem.

\begin{theorem}
\rm
\label{th2}
Unless W[1]=FPT, there does not exist an algorithm that can
decide whether a bipartite hamiltonian graph contains an induced matching of size $k$ or not in time $n^{o(k^\frac{1}{2})}$, where $n$ is the number of vertices in the graph, even when a hamiltonian cycle of the graph is available.

\begin{proof}
We assume there exists an algorithm $A$ that can determine whether a bipartite hamiltonian graph contains an induced matching of size $k$ or not in time $n^{o(k^{\frac{1}{2}})}$, where $n$ is the number of vertices in the graph.
Given an instance $(G,k)$ of the $(2k+1)$-{\sc Clique} problem, whose goal is to determine whether graph $G$ contains a clique of size $2k+1$ or not. We use the following algorithm to solve the $(2k+1)$-{\sc Clique} problem.
\begin{enumerate}
\item{Use the reduction we have developed in the proof of Theorem \ref{th1} to construct a bipartite hamiltonian graph $H$ from $(G, k)$ and compute a hamiltonian cycle $P$ in $H$;}
\item{apply $A$ to $H$ to decide whether it contains an induced matching of size $s=6k(2k+1)$ or not;}
\item{return ``yes'' if $A$ returns ``yes'', otherwise return ``no''.}
\end{enumerate}
From the reduction, we know that $H$ contains an induced matching of size $s$ if and only if $G$ contains a clique of size $2k+1$. The above algorithm thus correctly solves the $(2k+1)$-{\sc Clique} problem. Since $H$ contains $O(k^{2}n^{2})$ vertices, where $n$ is the number of vertices in $G$, the computation time needed by the algorithm to determine whether $H$ contains an induced matching of size $s$ is at most
\begin{equation}
n^{o(s^{\frac{1}{2}})}=n^{o(k)}
\end{equation}

The $(2k+1)$-{\sc Clique} problem can thus be solved
in time $n^{o(k)}$. However, it has been shown in Lemma \ref{lm1} that such an algorithm does not exist unless W[1]=FPT. Such an algorithm thus does not exist for the induced matching problem in hamiltonian bipartite graphs unless W[1]=FPT.
\end{proof}
\end{theorem}

\section{Inapproximability}
A well known inapproximability result regarding the {\sc Maximum Independent Set} problem was obtained in \cite{hastad}. It is shown that approximating the {\sc Maximum Independent Set} problem within a ratio of $n^{1-\epsilon}$ in polynomial time is NP-hard, where $n$ is the number of vertices in the graph and $\epsilon$ is any positive constant.

Based on this inapproximability result, we first show that it is NP-hard to approximate the maximum induced matching in general graphs within a ratio of $n^{1-\epsilon}$, where $n$ is the number of vertices in the graph and $\epsilon$ is any positive
constant. We then show that the same inapproximability result holds for the problem when the underlying graph is hamiltonian and a hamiltonian cycle of the graph is available. Finally, we show that unless NP=P,  the problem cannot be approximated within a ratio of $n^{\frac{1}{4}-\epsilon}$ when the underlying graph is hamiltonian bipartite and a hamiltonian cycle of the graph is available.

\begin{figure}[!tcp]
\begin{center}
\includegraphics[width=10.0cm, height=4.5cm]{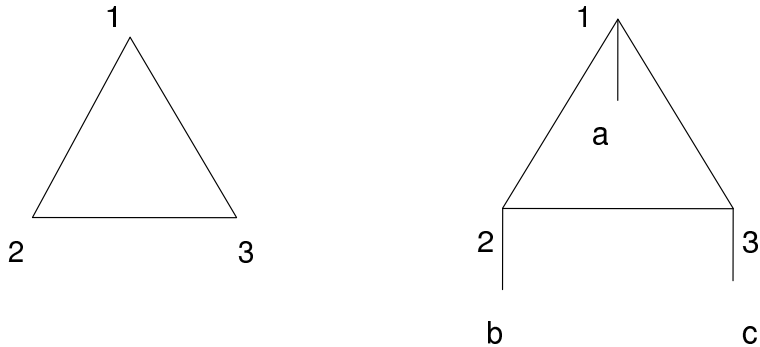}
\caption{A graph constructed by the reduction in the proof of Theorem \ref{th3}.}
\label{fig3}
\end{center}
\end{figure}

\begin{theorem}
\rm
\label{th3}
It is NP-hard to approximate a maximum induced matching
in a graph within a ratio of $n^{1-\epsilon}$ in polynomial time, where $n$ is the number of vertices in the graph
and $\epsilon$ is any positive constant.

\begin{proof}
We construct a simple reduction from the {\sc Maximum Independent Set} problem to the maximum induced matching problem. Given a graph $G=(V,E)$, the {\sc Maximum Independent Set} problem is to find a maximum independent set in $G$. For each vertex $u \in V$, we create an {\it image vertex} $i_{u}$ and connect $u$ and $i_u$ by an edge. This creates a new graph $H$.

Figure \ref{fig3} shows an example of the graph constructed by the reduction. The left part of the figure shows a graph $G$ that contains three vertices $1$, $2$ and $3$,
while its right part shows the graph $H$ constructed
from $G$ by connecting each vertex in $G$ into its image vertex; $a$, $b$, $c$ are the image
vertices of $1$, $2$ and $3$ respectively.

We assume there exists a polynomial time algorithm that can approximate the maximum induced matching in a graph within
a ratio of $n^{1-\epsilon}$. We apply the algorithm to
$H$ and obtain an induced matching $M$
in $H$. We denote the size of $M$ by $APX(H)$.
We then check every edge in $M$. If an edge of $M$ connects
a vertex in $G$ and its image vertex, we include
this vertex in a set $I$, otherwise we arbitrarily
select one of the two vertices in the edge and include it in $I$. Since $M$ is an induced matching, $I$ is an independent set. We denote the size of $I$ by $APX(G)$. The size of
$I$ is equal to the size of $M$. In other words, $APX(H)=APX(G)$.

We denote the size of the maximum independent set in $G$ with $OPT(G)$ and
the size of the maximum induced matching in $H$ with $OPT(H)$. We have
$OPT(G) \leq OPT(H)$ since given an independent set in $G$, we can immediately
obtain an induced matching in $H$ by selecting the edges that connect
vertices in the independent set to their image vertices in $H$. On the
other hand, we also have $OPT(G) \geq OPT(H)$ since an independent set
of size at least $OPT(H)$ can be obtained from an induced matching in
$H$ with the method we have described above. We thus have $OPT(G)=OPT(H)$.

Since the algorithm can approximate the maximum induced matching within a ratio of $n^{1-\epsilon}$, we have
\begin{equation}
\frac{APX(H)}{OPT(H)} \geq \frac{1}{(2|V|)^{1-\epsilon}} \geq \frac{1}{|V|^{1-\frac{\epsilon}{2}}}
\end{equation}
where the second inequality holds for sufficiently large $|V|$. However, since $APX(H)=APX(G)$ and $OPT(H)=OPT(G)$,
we immediately obtain
\begin{equation}
\frac{APX(G)}{OPT(G)} \geq \frac{1}{|V|^{1-\frac{\epsilon}{2}}}
\end{equation}
This suggests that there exists an algorithm
that can approximate a maximum independent set
in graph $G$ within a ratio of $n^{1-\frac{\epsilon}{2}}$
in polynomial time, which is contradictory to
the inapproximability result known for this problem. Such an
approximate algorithm thus does not exist for the maximum induced matching problem unless P=NP.
\end{proof}
\end{theorem}

\begin{theorem}
\rm
\label{th4}
It is NP-hard to approximate a maximum induced matching in a hamiltonian graph based on a hamiltonian cycle in it within a ratio of $n^{1-\epsilon}$ in polynomial time, where $n$ is the number of vertices in the graph and $\epsilon$ is any positive constant.

\begin{proof}
We construct a reduction from the maximum induced matching problem in a general graph to the same problem in a hamiltonian graph. Given a graph $G=(V,E)$, the goal of the maximum induced matching problem is to compute a maximum induced matching in $G$. We construct a graph $H$ based on $G$. We denote the number of vertices in $G$ by $p$ and create $p$ additional vertices $b_1, b_2, \cdots, b_p$. $b_1, b_2, \cdots, b_p$ are connected into a clique. In addition, each vertex in $G=(V,E)$ is joined to $b_1, b_2, \cdots, b_p$ by $p$ edges. We denote the resulting graph by $H$.

We first show that $H$ is hamiltonian. We use $u_1, u_2, \cdots, u_p$ to denote the vertices in $G$. From the construction of $H$, it is not difficult to see that $b_1, u_1, b_2, u_2, b_3, u_3, \cdots, b_{i}, u_{i}, \cdots, b_{p}, u_{p}, b_1$ form a hamiltonian cycle in $H$ and such a hamiltonian cycle can be obtained in polynomial time.

We use $OPT(G)$ to denote the size of a maximum induced matching in $G$ and $OPT(H)$ to denote the size of a maximum induced induced matching in $H$. We show that $OPT(G)=OPT(H)$. Since any induced matching $M$ in $G$ is also an induced matching in $H$, we immediately obtain $OPT(G) \leq OPT(H)$. On the other hand, we assume $N$ is a maximum induced  matching in $H$. From the construction of $H$, $N$ contains only one edge if $N$ contains an edge in the clique formed by $b_1, b_2, \cdots, b_p$, $OPT(H) \leq OPT(G)$ thus holds in this case. In addition, $OPT(H)=|N| \leq OPT(G)$ also holds if $N$ only contains edges in $G$. Finally, if $N$ contains an edge $e$ that joins a vertex $u$ in $G$ and one of the additional vertices $b_1, b_2, \cdots, b_p$, $N$ contains only one edge and $OPT(H)=|N| \leq OPT(G)$ again holds. We thus obtain $OPT(G)=OPT(H)$.

We now assume that there exists a polynomial time algorithm $A$ that can approximate the maximum induced matching in a hamiltonian graph based on a hamiltonian cycle in it within a ratio of $n^{1-\epsilon}$, where $n$ is the number of vertices in the graph and $\epsilon$ is some positive constant. Given a graph $G=(V,E)$, we use the following algorithm to approximate the maximum induced matching in $G$.
\begin{enumerate}
\item{Construct a hamiltonian graph $H$ from $G$ as described in the above reduction;}
\item{compute a hamiltonian cycle in $H$ as described above;}
\item{apply $A$ to $H$ to obtain an approximate solution $S$ for a maximum induced matching in $H$;}
\item{arbitrarily pick an edge $e$ in $G$ and return $e$ if $S$ contains only one edge;}
\item{otherwise return $S$.}
\end{enumerate}

It is clear that the algorithm returns an induced matching in $G$ in polynomial time. We use $APX(G)$ to denote the size of the matching returned by this algorithm and $APX(H)$ to denote the size of the matching returned by algorithm $A$. We then analyze the approximation ratio of this algorithm. First, it is straightforward to see that $APX(G)=APX(H)$. Since $H$ contains $2p$ vertices, we have
\begin{eqnarray}
\frac{OPT(G)}{APX(G)} & = & \frac{OPT(H)}{APX(H)} \\
                      & \leq & (2p)^{1-\epsilon} \\
                      & \leq & p^{1-\frac{\epsilon}{2}}
\end{eqnarray}
where the first inequality is due to the approximation ratio of $A$, and the last inequality holds for sufficiently large $p$.

$B$ thus can approximate a maximum induced matching in $G$ within a ratio of $p^{1-\frac{\epsilon}{2}}$ in polynomial time for sufficiently large $p$ and a positive constant $\epsilon$. This contradicts Theorem \ref{th3}, $A$ thus does not exist unless NP=P.
\end{proof}
\end{theorem}

\begin{theorem}
\rm
\label{th5}
It is NP-hard to approximate a maximum induced matching in a bipartite graph within a ratio of $n^{\frac{1}{4}-\epsilon}$ in polynomial time, where $n$ is the number of vertices in the graph and $\epsilon$ is any positive constant.

\begin{proof}
We construct a reduction from the {\sc Maximum Independent Set} problem to the maximum induced matching problem in bipartite graphs. Given a graph $G=(V,E)$, the goal of the {\sc Maximum Independent Set} problem is to compute a maximum independent set in $G$. We use $n$ to denote the number of vertices in $G$ and $u_1, u_2, \cdots, u_n$ to denote the vertices in $G$.

We assume that there exists a polynomial time algorithm $A$ that can approximate a maximum induced matching in a bipartite graph within a ratio of $N^{\frac{1}{4}-\epsilon}$, where $N$ is the number of vertices in the graph and $\epsilon$ is some positive constant. Without loss of generality, we assume that $\epsilon < \frac{1}{100}$.

For each vertex $u_i \in G$, where $ 1 \leq i \leq n$, we create two vertex groups, the two groups are included in the two sides of $H$. Each group contains $n^{3}$ vertices that represent $u_i$. We use $s_i$ and $t_i$ to denote the two groups for vertex $u_i$ in the two sides of $H$. $H$ thus contains $N=2n^{4}$ vertices in total.

For each $1 \leq i \leq n$, each vertex in $s_i$ is joined to only one vertex in $t_i$ by an edge in $H$. These edges are {\it homogeneous edges}. A pair of vertices $c \in s_i$, $d \in t_j$, where $i \neq j$, are joined by an edge in $H$ if $(u_i, u_j)$ is an edge in $G$. These edges are {\it heterogenous edges}.

We use $OPT(H)$ to denote the size of a maximum matching in $H$ and $OPT(G)=m$ to denote the size of a maximum independent set in $G$. Indeed, we assume $I=\{u_{i_1}, u_{i_2}, \cdots, u_{i_m}\}$ is a maximum independent set in $G$. For each $l \in \{i_1, i_2, \cdots, i_m\}$, all homogenous edges that join vertices in $s_{l}$, $t_{l}$ can be included into a matching $M'$. It is straightforward to see that since $I$ is an independent set, $M'$ is an induced matching of size $n^{3}OPT(G)$. This implies that $OPT(H) \geq n^{3}OPT(G)$.

We assume $M$ is a maximum induced matching in $H$. Since $M$ is a maximum induced matching, if a homogeneous edge that joins two vertices from $s_i$ and $t_i$, where $1 \leq i \leq n$, is included in $M$, all homogenous edges formed between vertices in $s_i$ and $t_i$ are also in $M$. If a heterogenous edge formed between $s_i$ and $t_j$, where $ i \neq j$, is included in $M$. Other heterogenous edges formed between $s_i$ and $t_j$ are not in $M$. The number of heterogeneous edges in $M$ is thus at most $n(n-1)$.

We consider the following algorithm $B$ for approximating a maximum independent set in $G$.
\begin{enumerate}
\item{construct a bipartite graph $H$ from $G$ based on the reduction described above;}
\item{apply $A$ to $H$ to find an approximate solution $S$ for a maximum induced matching in $H$;}
\item{we assume $S$ contains homogeneous edges formed between groups for vertices $h_1, h_2 \cdots, h_l$ in $G$, return $\{h_1, h_2, \cdots, h_l\}$ as an approximate solution for a maximum independent set in $G$.}
\end{enumerate}

It is straightforward to see that $B$ returns an independent set in $G$ in polynomial time. We use $APX(H)$ to denote the size of the induced matching returned by $A$ and $APX(G)$ to denote the size of the independent set returned by $B$. We first show that $APX(G) \geq 1$.  If $B$ returns an empty set, $S$ only contains heterogenous edges. This implies that $APX(H) < n^{2}$. We thus have
\begin{eqnarray}
\frac{OPT(H)}{APX(H)} & > & \frac{n^{3}OPT(G)}{n^{2}} \\
                      & \geq & n \\
                      & = & (\frac{N}{2})^{\frac{1}{4}} \\
                      & > &  N^{\frac{1}{4}-\epsilon}
\end{eqnarray}
where the first inequality follows from $OPT(H) \geq n^{3}OPT(G)$ and $APX(H) < n^{2}$; the last inequality holds for sufficiently large $N$. This contradicts the approximation ratio of $A$. $B$ thus at least returns one vertex in $G$ and $APX(G) \geq 1$. From the approximation ratio of $A$, we have
\begin{eqnarray}
\frac{OPT(H)}{APX(H)} & \leq & N^{\frac{1}{4}-\epsilon} \\
                      &  = & (2n^{4})^{\frac{1}{4}-\epsilon}
\end{eqnarray}
Since $OPT(H) \geq n^{3}OPT(G)$ and $APX(H) \leq n^{3}APX(G)+n^{2}$, we have
\begin{equation}
\frac{n^{3}OPT(G)}{n^{3}APX(G)+n^2} \leq \frac{OPT(H)}{APX(H)}
\end{equation}
Since $APX(G) \geq 1$, we have
\begin{equation}
\frac{n^{3}OPT(G)}{n^{3}APX(G)+n^{2}} \geq \frac{OPT(G)}{2APX(G)}
\end{equation}
This immediately leads to
\begin{eqnarray}
\frac{OPT(G)}{APX(G)} &  \leq & 2(2n^{4})^{\frac{1}{4}-\epsilon} \\
                      & \leq & n^{1-\epsilon}
\end{eqnarray}
where the last inequality holds for sufficiently large $n$.

When the number of vertices in $G$ is sufficiently large, $B$ can approximate the maximum independent set in $G$ within a ratio of $n^{1-\epsilon}$ in polynomial time, where $n$ is the number of vertices in $G$ and $\epsilon$ is some positive constant. This contradicts the inapproximability of the {\sc Maximum Independent Set} problem. $A$ thus does not exist for the maximum induced matching problem in bipartite graphs.
\end{proof}
\end{theorem}

A bipartite graph is {\it equally sided} if its two sides contain the same number of vertices. Since the bipartite graph $H$ constructed in the proof of Theorem \ref{th5} is equally sided. We immediately obtain the following Corollary.

\begin{corollary}
\rm
\label{col1}
It is NP-hard to approximate a maximum induced matching in a equally sided bipartite graph within a ratio of $n^{\frac{1}{4}-\epsilon}$, where $n$ is the number of vertices in the graph and $\epsilon$ is any positive constant.
\end{corollary}

\begin{theorem}
\rm
\label{th6}
It is NP-hard to approximate a maximum induced matching in a hamiltonian bipartite graph based on a hamiltonian cycle of the graph within a ratio of $n^{\frac{1}{4}-\epsilon}$ in polynomial time, where $n$ is the number of vertices in the graph and $\epsilon$ is any positive constant.

\begin{proof}
We reduce the maximum induced matching problem in an equally sided bipartite graph to the same problem in a hamiltonian bipartite graph. Given an equally sided bipartite graph $G=(V_{1}\cup V_{2}, E)$, where $|V_1|=|V_2|$, the goal of the problem is to compute a maximum induced matching in $G$.

We construct a hamiltonian bipartite graph based on $G$. Specifically, we denote  $p=|V_1|=|V_2|$ and create $2(p+1)$ additional vertices $l_1, l_2, \cdots, l_{p+1}$ and $m_1, m_2, \cdots, m_{p+1}$. These additional vertices are connected into a complete bipartite graph such that $l_1, l_2, \cdots, l_{p+1}$ are in one side and $m_1, m_2, \cdots, m_{p+1}$ are in the other side.

Each vertex in $V_1$ is joined to all of $m_1, m_2, \cdots, m_{p+1}$ by $p+1$ edges. Similarly, each vertex in $V_2$ is joined to all of $l_1, l_2, \cdots, l_{p+1}$ by $p+1$ edges. We denote the resulting bipartite graph by $H$.

First, we show $H$ is hamiltonian. We use $u_1, u_2, \cdots, u_{p}$ to denote the vertices in $V_1$ and $v_1, v_2, \cdots, v_p$ to denote the vertices in $V_2$. From the construction of $H$, it is straightforward to see that $m_1, u_1, m_2, u_2, \cdots, m_{p}, u_{p}, m_{p+1}, l_1, v_1, l_2, v_2, \cdots, l_{p}, v_{p}, l_{p+1}, m_{1}$ form a hamiltonian cycle in $H$. Such a cycle can be obtained in polynomial time.

We use $OPT(H)$ to denote the size of a maximum induced matching in $H$ and $OPT(G)$ to denote the size of a maximum induced matching in $G$. Since an induced matching in $G$ is also an induced matching in $H$, we immediately obtain $OPT(G) \leq OPT(H)$. We assume $M$ is a maximum induced matching in $H$ and $|M| \leq OPT(G)$ if $M$ only contains edges in $G$; $|M|=1$ if $M$ contains an additional vertex. This implies that $OPT(H) \leq OPT(G)$. We thus have $OPT(H)=OPT(G)$.

We assume there exists a polynomial time algorithm $A$ that can approximate a maximum induced matching in a hamiltonian bipartite graph. We consider the following algorithm $B$ for approximating a maximum induced matching in an equally sided bipartite graph $G$.
\begin{enumerate}
\item{Construct a hamiltonian bipartite graph $H$ as described above;}
\item{compute a hamiltonian cycle $P$ in $H$;}
\item{apply $A$ to $H$ to obtain an approximate solution $S$ for a maximum induced matching in $H$;}
\item{return $S$ if it only contains edges in $G$;}
\item{otherwise, arbitrarily choose an edge $e$ in $G$ and return $e$.}
\end{enumerate}

It is clear that $B$ returns an induced matching in $G$ in polynomial time. We use $APX(G)$ to denote the size of the induced matching returned by $B$ and $APX(H)$ to denote the size of the induced matching returned by $A$. It is straightforward to see that $APX(G)=APX(H)$. We thus have
\begin{eqnarray}
\frac{OPT(G)}{APX(G)} & = & \frac{OPT(H)}{APX(H)} \\
                      & \leq & (4q+1)^{\frac{1}{4}-\epsilon} \\
                      & \leq & (2q)^{\frac{1}{4}-\frac{\epsilon}{2}}
\end{eqnarray}
where the first inequality is due to the approximation ratio of $A$ and the last inequality holds for sufficiently large $q$.

$B$ thus can approximate a maximum induced matching in an equal sided bipartite graph within a ratio of $n^{\frac{1}{4}-\frac{\epsilon}{2}}$ in polynomial time, where $n$ is the number of vertices in the graph and $\epsilon$ is some positive constant. This contradicts Corollary \ref{col1}. $A$ thus does not exist and the theorem has been proved.
\end{proof}
\end{theorem}

\section{Conclusions}

In this paper, we study the parameterized complexity and inapproximability of the {\sc Induced Matching} problem in hamiltonian bipartite graphs. Our results show that although hamiltonian bipartite graphs contain some additional structure features when compared with general graphs, these structure features cannot lead to efficient parameterized algorithm for this problem. Our work also establishes inapproximablity results for computing a maximum induced matching in hamiltonian bipartite graphs.

Although we have shown that a maximum induced matching in a hamiltonian bipartite graph cannot be approximated within a ratio of $n^{\frac{1}{4}-\epsilon}$ in polynomial time, where $n$ is the number of vertices in the graph and $\epsilon$ is any positive constant, it remains open whether there indeed exists an approximation algorithm that can in fact achieve an approximation ratio close to it. We believe such an algorithm exists for the {\sc Maximum Induced Matching} problem in hamiltonian bipartite graphs and our future work may focus on the development and analysis of such an algorithm.

\section*{Acknowledgments}
The author's work is fully supported by the University Fund of Jiangsu University of Science and Technology under the grant number: 635301202.


\begin{thebibliography}{25}
\bibliographystyle{plain}
\bibitem{baker}
Brenda S. Baker, ``Approximation Algorithms for NP-complete Problems on Planar Graphs'', {\it Journal of ACM},
41(1): 153-180, 1994.
\bibitem{biggs}
N.L. Biggs, E.K. Lioyd, and R.J. Wilson, {\it Graph Theory}, Clarendoh Press, Oxford, 8-9, 1976.
\bibitem{Boppana}
Ravi Boppana and Magnus M. Halldorsen, ``Approximating Maximum Independent Set by Excluding Subgraphs'', {\it BIT Numerical Mathematics}, 32(2):180-196, 1992
\bibitem{cameron1}
Kathic Cameron, ``Induced Matchings in Intersection Graphs'', {\it Discrete Mathematics}, 278: 1-9, 2003.
\bibitem{cameron2}
Kathic Cameron, R. Sritharan and Yiwen Tang, ``Finding a Maximum Induced Matching in Weakly Chordal Graphs'', {\it Discrete Mathematics}, 266(1-3): 133-142, 2003.
\bibitem{chen}
Jianer Chen, Xiuzhen Huang, Iyad A. Kanj, Ge Xia, ``Linear FPT reductions and
computational lower bounds'', {\it Proceedings of the Thirty-Sixth ACM Symposium on Theory of Computing (STOC 2004)},
212-221, 2004.
\bibitem{chen0}
Jianer Chen, Iyad A. Kanj, Ge Xia, ``Improved Parameterized Upper Bounds for Vertex Cover'', {\it Proceedings of the Thirty-First International Symposium on Mathematical Foundations of Computer Science(MFCS 2006)}, 238-249, 2006.
\bibitem{child}
Andrew M. Childs, Quantumn information processing in continuous time, {\it Ph.D. Dissertation}, Massachusetts Institute of Technology, June 2004.
\bibitem{dinur}
Irit Dinur and Shmuel Safra, ``The Importance of Being Biased'', {\it Proceeding of the Thirty-Fourth ACM Symposium on
Theory of Computing (STOC 2002)}, 33-42, 2002.
\bibitem{downey}
Rodney G. Downey and Michael R. Fellows, {\it Parameterized Complexity}, Springer-Verlag, 1998.
\bibitem{downey1}
Rodney G. Downey and Michael R. Fellows, ``Fixed Parameter Tractibility and Completeness i: Basic Theory'',
{\it SIAM Journal of Computing}, 24:873-921, 1995.
\bibitem{downey2}
Rodney G. Downey and Michael R. Fellows, ``Fixed Parameter Tractibility and Completeness ii: Completeness for W[1]'',
{\it Theoretical Computer Science A}, 141:109-131, 1995.
\bibitem{downey3}
Rodney G. Downey, Richanel R. Fellows, and Venkatesh Raman, ``The complexity of irredundant set parameterized by size'',
{\it Discrete Applied Mathematics}, 100(3): 155-167, 2000.
\bibitem{duckworth}
William Duckworth, David F. Manlove, and Michele Zito, ``On the Approximability of the Maximum Induced Matching Problem'', {\it Journal of Discrete Algorithms}, 3(1):79-91, 2005.
\bibitem{fleury}
M. Fleury, ``Deux probl\`{e}mes de g\'{e}om\'{e}trie de situation'', {\it Journal de Math\'{e}matiques \'{E}l\'{e}mentaires, 2nd ser (in French)}, 2: 257-261, 1883.
\bibitem{galil}
Zvil Galil, ``Efficient algorithms for finding maximum matching in graphs'', {\it ACM Computing Surveys (CSUR)}, 18(1): 23-28, 1986.
\bibitem{garey}
M. R. Garey and D. S. Johnson, {\it Computers and Intractibility}, W. H. Freeman and Co., San Francisco, California, 1979. A
guide to the theory of NP-completeness, A Series of Books in the Mathematical Sciences.
\bibitem{hastad}
Johan Hastad, ``Clique is hard to approximate within $n^{1-\epsilon}$'', {\it Proceedings of the Thirty-Seventh Annual Symposium on Foundations of
Computer Science (FOCS 1996)}, 627-636, 1996.
\bibitem{Johnson}
David S. Johnson, ``Approximate Algorithms for Combinatorial Problems'', {\it Journal of Computer and System Sciences}, 9, 256-278, 1974.
\bibitem{kanj}
Iyad Kanj, Michael J. Pelsmajer, Marcus Schaefer, and Ge Xia, ``On the induced matching problem", {\it Proceedings of the Twenty-Fifth Symposium on Theoretical Aspects of Computer Science (STACS 2008)}, 397-408, 2008.
\bibitem{kao}
Shin-Shin Kao and Lih-Hsing Hsu,``Spider web networks: a family of optimal, fault tolerant, hamiltonian bipartite graphs'', {\it Applied Mathematics and Computation}, 160: 269-282, 2005.
\bibitem{liu0}
C. Liu and Y. Song, ``Parameterized dominating set problem in chordal graphs: complexity and lower bound'', {\it Journal of Combinatorial Optimization}, 18(1): 87-97, 2009.
\bibitem{liu}
C. Liu and Y. Song, ``Parameterized complexity and inapproximability of dominating set problem in chordal and near chordal graphs'', {\it Journal of Combinatorial Optimization}, 22: 684-698, 2011.
\bibitem{moser1}
Hannes Moser and Somnath Sikdar, ``The parameterized complexity of the induced matching in planar graphs'', {\it Proceedings of the First International Frontiers of Algorithmics Workshop (FAW 2007)}, 325-336, 2007.
\bibitem{moser2}
Hannes Moser and Dimitrios M. Thilikos, ``The Parameterized complexity of the induced matching problem'', {\it Discrete Applied Mathematics}, 157:715-727, 2009.
\bibitem{raz}
Ran Raz and Shmuel Safra, ``A Sub-Constant Error-Probability Low Degree Test, and a Sub-Constant
Error-Probability PCP Characterization of NP'', {\it Proceedings of the Twenty-Ninth ACM Symposium on Theory of Computing (STOC 1997)},
475-484, 1997.
\end{thebibliography}
\end{document}